\documentclass[useAMS,usenatbib,usegraphicx]{mn2e}
\usepackage{epsfig}
\usepackage{times} %% For PDF
\usepackage{amssymb,amsmath}

\def\mum{$\, \mu$m}
\def\mums{$\, \mu$m\ }

\def\arcs{\hbox{$^{\prime\prime}$}}

\begin{document}

%TITLES AND AUTHORS
\title[Measurement of the Sunyaev-Zel'dovich Increment in massive
galaxy clusters]
{Measurement of the Sunyaev-Zel'dovich Increment in massive galaxy clusters}

\author[Zemcov et al.]{
\parbox[t]{\textwidth}{
\vspace{-1.0cm}
Michael Zemcov$^{1}$, 
Mark Halpern$^{1}$,
Colin Borys$^{2}$,
Scott Chapman$^{2}$,
Wayne Holland$^{3}$,
Elena Pierpaoli$^{4}$,
Douglas Scott$^{1}$
}
\vspace*{6pt}\\
$^{1}$ Department of Physics \& Astronomy, University of British Columbia,
Vancouver, BC V6T 1Z1, Canada \\
$^{2}$ Division of Physics, Math \& Astronomy, California Institute of
Technology, Pasedena, CA 91125, USA \\
$^{3}$ UK Astronomy Technology Centre, Royal Observatory, Blackford
Hill, Edinburgh EH9 3HJ, UK \\
$^{4}$ Physics Department and Astronomy Department, Princeton
University, NJ 08540, USA \\
\vspace*{-0.5cm}}

\date{Draft version \today}

\maketitle

\begin{abstract}
We have detected the Sunyaev-Zel'dovich (SZ) increment at $850$\mums
in two galaxy clusters (Cl$\, 0016+16$ and MS$\, 1054.4-0321$) using
SCUBA (Sub-millimetre Common User Bolometer Array) on the James Clerk
Maxwell Telescope.  Fits to the isothermal $\beta$ model yield a
central Compton $y$ parameter of $(2.2 \pm 0.7) \times 10^{-4}$ and
a central $850$\mums flux of $\Delta I_{0} = 2.2 \pm 0.7
\,$mJy beam$^{-1}$ in Cl$\, 0016$.  This can be combined with
decrement measurements to infer $y = (2.38 \pm _{0.34}^{0.36}) \times
10^{-4}$ and $v_{\mathrm{pec}} = 400 \pm _{1400}^{1900} \,$km s$^{-1}$.
In MS$\, 1054$ we find a peak $850$\mums flux of $\Delta I_{0} = 2.0
\pm 1.0 \,$mJy~beam$^{-1}$ and $y = (2.0 \pm 1.0) \times 10^{-4}$.
To be successful such measurements require large chop throws and
non-standard data analysis techniques.  In particular, the $450$\mums
data are used to remove atmospheric variations in the $850$\mums data.
An explicit annular model is fit to the SCUBA difference data in order
to extract the radial profile, and separately fit to the model
differences to minimize the effect of correlations induced by our
scanning strategy.  We have demonstrated that with sufficient care,
SCUBA can be used to measure the SZ increment in massive, compact
galaxy clusters.
\end{abstract}

\begin{keywords}
cosmic microwave background -- cosmology: observations -- galaxies:
clusters: general -- large-scale structure of universe -- methods:
observational -- submillimetre
\vspace*{-0.5cm}
\end{keywords}

\section{Introduction}

The Sunyaev-Zel'dovich (SZ) effect can be used as a powerful probe of
cosmology.  Because the intensity of the SZ distortion is virtually
independent of redshift, SZ measurements provide an avenue to study
clusters and their peculiar velocities to any distance.  Additionally,
the SZ effect allows independent determination of various cosmological
parameters.  For example, because the SZ effect's amplitude is
proportional to the electron density along the line of sight through
the cluster, $n_{\mathrm{e}}$, while the X-ray amplitude is
proportional to $n_{\mathrm{e}}^{2}$, SZ effect data can yield
distances to clusters independently from the cosmological distance
ladder.  This information can then be used to provide estimates of the
Hubble constant $H_{0}$ (e.g. \citealt{Reese2000}).

The physics of the SZ effect is quite simple (\citealt{SZ1972},
\citealt{Birkinshaw1999}, \citealt{Carlstrom2002}). Cosmic microwave
background (CMB) photons can scatter off hot ($T_{\mathrm{e}} \simeq
10^{7} \,$K) electrons in a plasma.  A statistical net gain of energy
of the photons from the electrons occurs, producing a characteristic
distortion in the CMB as seen through the plasma.  This distortion
produces a \textit{decrement} in the CMB's temperature below about 200
GHz, but an \textit{increment} in the CMB's temperature above this
frequency.

The most important use of SZ increment measurements lies in the
determination of the spectral shape of the SZ effect in clusters of
galaxies.  This shape is the sum of the thermal and kinetic effects
together with other sources of emission within the cluster.  The
kinetic SZ effect is a change in the apparent CMB temperature as seen
through the cluster due to the peculiar motion of the cluster relative
to the CMB.  The kinetic SZ effect may be present in any given cluster
and may manifest itself as either a temperature increase or decrease,
although it never dominates the thermal effect for expected cluster
velocities \citep{Birkinshaw1999}.  Knowledge of the cluster's
properties obtained at several wavelengths can allow separation of
these two effects.  If the kinetic effect can be isolated, the
cluster's peculiar velocity along the line of sight can be determined.
Statistics about large scale flows at a variety of redshifts provide
strong constraints on the dynamics of structure formation.  In
principle, detailed measurement of the SZ spectral shape can also
constrain the cluster gas temperature through the relativistic
correction (\citealt{Rephaeli1995}, \citealt{Birkinshaw1999},
\citealt{Colafrancesco2003}).

High frequency measurements can also help separate contaminants such
as dusty galaxies which may exist in millimetric measurements, since
the dominant sub-mm point sources differ from those seen at longer
wavelengths.

Measurement of the Sunyaev-Zel'dovich (SZ) effect increment in galaxy
clusters has been an elusive and important goal of observational
cosmology for over a decade. While measurements of the SZ effect
decrement are common (\citealt{Birkinshaw1999},
\citealt{Carlstrom2002}), measurement and study of the increment has
been less successful. This is largely because current instruments
generally lack the sensitivity to measure small amplitude, extended,
positive signals in the sky.  Also, other positive flux sources in a
cluster field can confuse SZ measurements (\citealt{Loeb1997},
\citealt{Blain1998}).  However, experiments working near the null of
the thermal SZ effect like SuZIE \citep{Holzapfel1997}, and ACBAR
\citep{Peterson2002} are beginning to yield results.  The
balloon-borne experiment PRONAOS has had limited success in SZ
increment measurements \citep{Lamarre1998}, and other detections are
also claimed (e.g. \citealt{Komatsu1999}).  Recently,
\citet{Benson2003} have performed this type of measurement in six
galaxy clusters, and find that measurement of any given cluster's
peculiar velocity is difficult.

Measurement of the SZ increment is difficult for a variety of reasons,
including the amplitude of the atmospheric signal at sub-mm
wavelengths. $450$\mums and $850$\mums observations take advantage of
windows in the opacity of the sky at these wavelengths, but successful
observation of any astronomical source in the sub-mm regime still
requires non-trivial atmospheric subtraction techniques.
Additionally, because the SZ effect is a small amplitude signal
extended across a relatively large region of the sky, sensitive
instruments which can sample a range of spatial frequencies are
required.  Care must be taken when reconstructing the spatial shape of
the SZ effect from data recorded as differences between two or more
telescope pointings.  As point sources are expected to be a major
contaminant in the sub-mm regime, high resolution is also required to
adequately understand their effect in a given field.  One of the
instruments currently most capable of performing observations of the
SZ increment is the Sub-millimetre Common User Bolometer Array (SCUBA;
\citealt{Holland1999}).  This instrument provides enough sensitivity
and angular resolution to detect and remove point sources (including
lensed background sources) from the data.  SCUBA also samples large
enough regions of the sky to constrain the shape of the SZ increment
in moderate to high redshift clusters.  However, even with these
characteristics, measurement of the SZ increment with SCUBA is not
trivial.

\section{Observations}

The bulk of the data discussed here was obtained on November $10$ and
$11$, $1999$ at the James Clerk Maxwell Telescope (JCMT) at the summit
of Mauna Kea, Hawaii. The JCMT combines a high, dry site with a $15$
metre dish and sensitive instruments.  SCUBA, attached on the Nasmyth
platform of the JCMT, allows simultaneous observations with $91$
$450$\mum--band bolometers and $37$ $850$\mum--band bolometers.
Because the $450$\mums band observes very little SZ increment flux,
while the expected peak of the SZ signal occurs near $850$\mum, the
$450$\mums data can act as a monitor of atmospheric emission in this
experiment.

SCUBA is sparsely sampled spatially, which is undesirable in this
experiment.  `Jiggling' to fully sample the array consists of small
($\simeq 6$ arcsec) changes in the position of the beam in a regular
pattern.  The detectors integrate in each jiggle position for $1.024
\,$s.  Although $64$ positions are required to fully sample both
arrays, this experiment used $16$ positions because it was only
necessary to fully sample the $850$\mums array.  Moreover, executing
the jiggle pattern as quickly as possible is more conservative in
terms of residual sky fluctuations.

Variations in the signal produced by atmospheric (sky) noise and
thermal offsets from the instrument are removed by rapid chopping
together with slower telescope nodding (e.g. \citealt{Archibald2002}).
Rapidly chopping the secondary mirror faster than the rate at which
the sky is varying removes much of this atmospheric signal. Nodding
the telescope's primary mirror to a second reference position removes
slow sky noise gradients and instrument noise. These motions produce a
chop pattern such that the two reference beams are each observed half
as often as the source beam (Fig.~1). In the standard SCUBA mapping
mode, the chop occurs at about $8$ Hz and the nod occurs at about
$1/16$ Hz.  Although a chop throw in the range $50$ to $100$ arcsec is
standard, this experiment requires a chop that does not put
significant SZ flux in the reference beams (Fig.~1). For this reason,
an extremely large JCMT secondary mirror chop size of $180$ arcsec was
used.

\begin{figure}
\centering
\epsfig{file=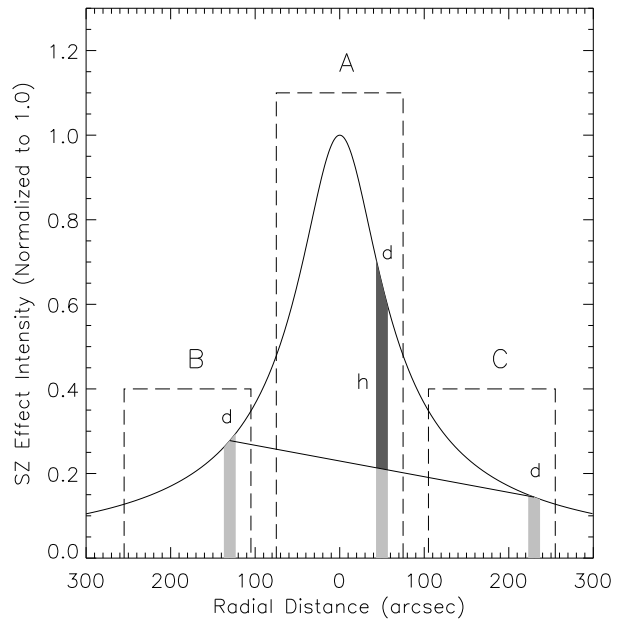,width=0.45\textwidth}
\caption{Our chopping strategy in relation to the SZ increment.  The
solid curve is an isothermal $\beta$ profile (discussed in Section
$3.3$) with parameters $\theta_{\mathrm{c}} = 50$ arcsec and $\beta =
0.75$.  The profile's peak has been normalized such that $\Delta
I_{0}= 1.0$.  The dashed lines show the SCUBA array size at the
`source' position marked `A', and the two reference positions, `B' and
`C'.  The filled areas marked `d' show the FWHM of one bolometer, and
indicate the bolometer's position over an observation.  SCUBA jiggle
mapping gives data in the form $h = d_{\mathrm{A}} - (d_{\mathrm{B}} +
d_{\mathrm{C}}) / 2$ for each bolometer.  Because it is sensitive only
to differences in SZ flux between the source and two reference beams,
$h$ contains much of the information about the SZ effect amplitude,
although this is the same information contained in the shape within
the array size.  However, the JCMT does not have high enough resolution
to rely on the differenced SZ shape alone.  Using the $850 \,$\mums
array average (that is, the mean of every bolometer's $h$) to remove
the atmospheric emission at each time step subtracts a flux about the
same size as $h$, which destroys the information about the SZ effect
amplitude in this data.}
\label{fig:1}
\end{figure}

$89 \,$ks of data were taken, split between several fields.  The CSO
$225 \,$GHz $\tau$-meter was not operational for either observing
shift, so we relied on skydips to measure the atmospheric optical
depth, $\tau_{850}$, in our data set. Both nights had relatively
constant $\tau_{850}$, but the first evening had a lower average
($\tau_{850} \simeq 0.15$) than the second ($\tau_{850} \simeq 0.3$).
Uranus was used as a calibration source throughout.

Three clusters and three blank fields were observed during this run
(see Table 1 for details), referred to as Cl$\, 0016$, MS$\, 0451$,
MS$\, 1054$, Blank $02$, Blank $09$ and Blank $25$ hereafter.  $23
\,$ks of additional data were collected with an aluminum reflector
masking the aperture of SCUBA. These data are used as null tests to
study SCUBA's behaviour under both optical and electrical influences
(blank sky data) and electrical influences only (reflector data).

Additional data were taken at the JCMT on October $2$, $3$, and $6$,
$2002$ in the same manner as the $1999$ run.  Both Cl$\, 0016$ and
another blank field, Blank $21$, were mapped. In addition to the
jiggle mapping, photometry of possible point sources in the Cl$\,
0016$ field were performed.  The photometry used a $9$ point jiggle
pattern with $1$ arcsec spacing and a $60$ arcsec chop throw.
Bolometer G9 of the SCUBA array was not operating properly during this
run and its data are not used here.  A total of $13 \,$ks of data were
taken in 2002; how they are combined with the earlier data is
discussed below.

\section{General Analysis}

\subsection{Data Preparation}

The data are deswitched, flat-fielded and corrected for atmospheric
extinction with {\sc surf} (the SCUBA User Reduction Facility,
\citealt{J&L1998}).  Extinction correction is performed by calculating
the airmass at which each bolometer measurement was made, and then
multiplying by the zenith sky extinction at the time of the
measurement to give the extinction optical depth along the line of
sight.  Each data point is then multiplied by the exponential of the
optical depth to give the value that would have been measured in the
absence of the atmosphere.  The photometric calibration of deep SCUBA
fields like those presented here is known to be accurate to about 15
per cent (\citealt{Borys2002a}, \citealt{Archibald2002}), although
$\sim 2/3$ of this uncertainty is due to error in the absolute
calibration of the planets' flux.

It is well documented that the deswitched $450$\mums and $850$\mums
time streams are highly correlated, and that this correlation stems
primarily from atmospheric noise (e.g. \citealt{JLH1998},
\citealt{Borys1999}, \citealt{Archibald2002}).  The standard method of
removing atmospheric noise from an array is to subtract the array
average from each bolometer at each time step.  However, most of the
information about the SZ flux in this measurement is contained in the
difference between the source and reference beam fluxes (see Fig.~$1$)
rather than in the shape of the SZ emission across the field of view.
Therefore, removing the $850 \,$\mums array average at each time step
would drastically reduce the inferred SZ emission.  However, because
the SZ effect is small at $450 \,$\mum, using that band's data should
mitigate this problem.  A least-squares linear fit of the $450$\mums
array average to the $850$\mums array average at each time step is
used to obtain a linear coefficient $C_{1}$ and an offset $C_{0}$
(hereafter referred to as the $\left \langle 450 \, \mu \mathrm{m}
\right \rangle$ method).  These fit parameters and the $450$\mums
array averages are then used in place of the $850$\mums array average
to subtract the atmospheric noise at each time:
\begin{equation}
S_{b}^{850}(t) = R_{b}^{850}(t) - \left ( C_{1} \left \langle
R^{450}(t) \right \rangle + C_{0} \right ),
\end{equation}
where $S_{b}^{850}(t)$ is the corrected data at time step $t$ for
bolometer $b$, $R_{b}^{850}(t)$ is the raw, double-differenced and
extinction corrected data at time step $t$ for bolometer $b$, the
angled brackets denote an average over the whole array, and the
superscripts denote wavelength.  The fits are performed separately for
each (approximately $20$ minute) observation.  $C_{1}$ removes the
atmospheric emission on scales of about $1 \,$s, while $C_{0}$ removes
the atmospheric average over periods long compared to the telescope
nod time.  The value of $C_{1}$ and $C_{0}$ do not vary by more than
10 per cent over an observation; this is consistent with the results of
\citet{Borys1999}.  $C_{1}$ is typically $0.6$ and $C_{0}$ is of order
$1 \, \mu$V in our data set.  The linear Pearson correlation
coefficients of the $450$\mums and the $850$\mums time streams are on
average $0.95$ and never less than $0.90$ for these observations.  The
average ratio of the variance of $S_{b}^{850}$ in the time stream of
the $\left \langle 450 \, \mu \mathrm{m} \right \rangle$ atmospheric
subtraction method to the variance of the $850 \,$\mums time stream in
the standard {\sc surf} atmospheric subtraction method is $1.15$.
This implies that the $\left \langle 450 \, \mu \mathrm{m} \right
\rangle$ method is slightly inferior for removing atmospheric noise.
However, this approach is required for the analysis of SZ data, since
it allows retention of information on angular scales up to that of the
chop throw.

\begin{table*}
\caption{Experimental target field parameters, including field names
and coordinates, on source integration times, and dates of
observation.}
\label{tab:1}
\begin{tabular}{llllll}
\hline 

Target	      & Field Type & RA (2000)       & DEC (2000)
                        & Integration & Observation \\
              &             &                       &                      
                        & Time (ks)   & Dates       \\ \hline

Blank $02$       & Blank Field   & 2:22:59.9  & +4:59:57   & $2.5$
& 11 Nov.~1999 \\

Blank $09$       & Blank Field   & 9:00:00.0  & +10:02:00  & $9.4$
& 10,11 Nov.~1999 \\

Blank $21$       & Blank Field   & 21:30:00.0 & +16:30:00  & $7.2$
& 03,06 Oct.~2002 \\

Blank $22$       & Blank Field   & 22:17:25.1 & +0:12:59   & $8.8$
& 11 Nov.~1999 \\

Cl$\, 0016$+$16$ & Cluster Field & 0:18:33.7  & +16:26:04  & $12.2$
& 11 Nov.~1999, 02,03 Oct.~2002 \\

MS$\, 0451$-$03$ & Cluster Field & 4:54:11.5  & $-3$:00:52 & $15.1$
& 10,11 Nov.~1999 \\

MS$\, 1054$-$03$ & Cluster Field & 10:56:57.4 & $-3$:37:24 & $15.1$
& 10,11 Nov.~1999 \\ \hline
\end{tabular}
\end{table*}

Of the seven target fields, three are unusable for this experiment in
some way. A fundamental requirement of this analysis method is high
enough signal to noise ratio that it is possible to determine the
location of point sources in a field and remove them. However, the
Blank $02$ field had only a short integration time (see Table 1),
hence it is not included in the results of this experiment.  Blank
$22$ was a poor choice for a control field, because it contains large
point sources which corrupt it for our purposes; \citet{Chapman2001}
examine this field in detail.  The MS$\, 0451$ field contains a sub-mm
bright gravitationally lensed arc serendipitously discovered when
combining this data with previous observations \citep{Chapman2002}.
Although an interesting target in its own right, the $\simeq 12 \,$mJy
beam$^{-1}$ resolved arc dominates the sub-mm emission in this cluster
\citep{Borys2003}, making measurement of the SZ increment impossible
with these data.

\subsection{Point Source Removal}

Recognition and removal of point sources, especially unbiased handling
of lensed background sources, is important in this experiment
(\citealt{Loeb1997}, \citealt{Blain1998}).  We adopt the following
procedure.  The measured flux differences are binned according to the
spatial position of the `source' beam.  The standard method of
presenting a SCUBA image is to convolve the binned data set with the
telescope point spread function (PSF).  Maps made by this method are
shown in Fig.~$2$ in order to indicate the locations of possible point
sources and to exhibit their consistency with the same fields in
\citet{Chapman2002}.  However, these images are not used as part of
the quantitative analysis.

\begin{figure*}
\centering
\epsfig{file=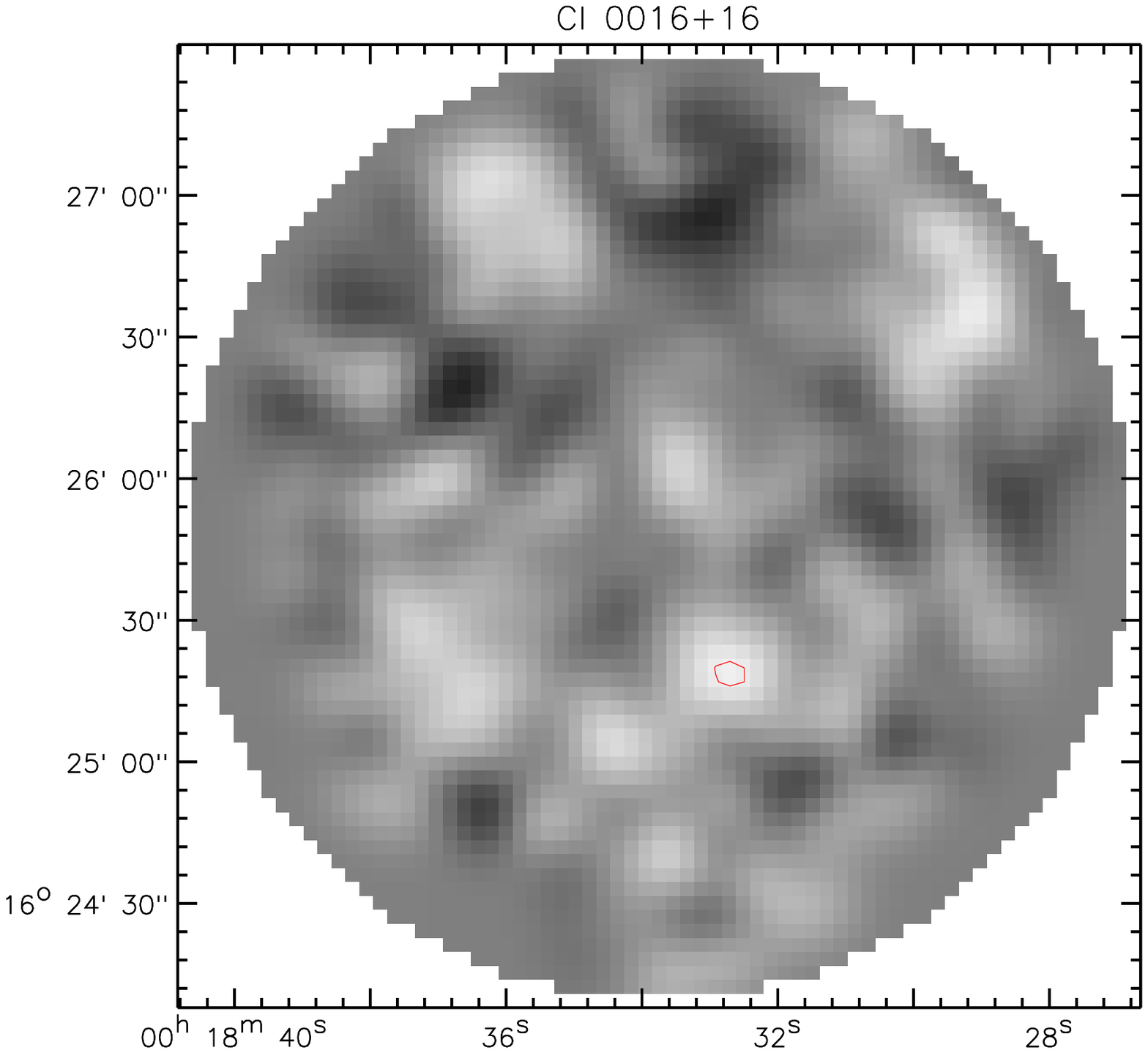,width=0.45\textwidth}
\epsfig{file=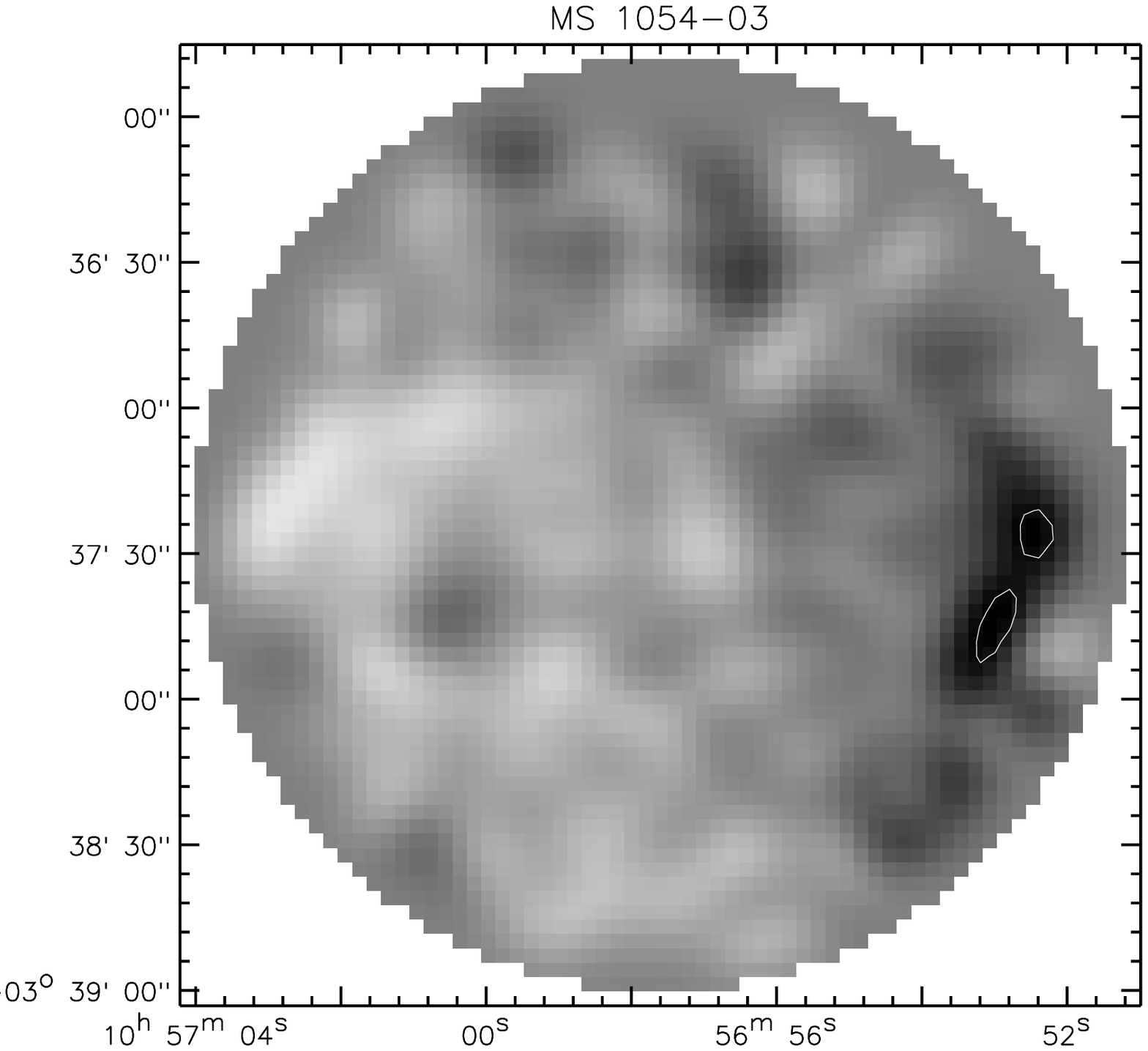,width=0.45\textwidth}
\epsfig{file=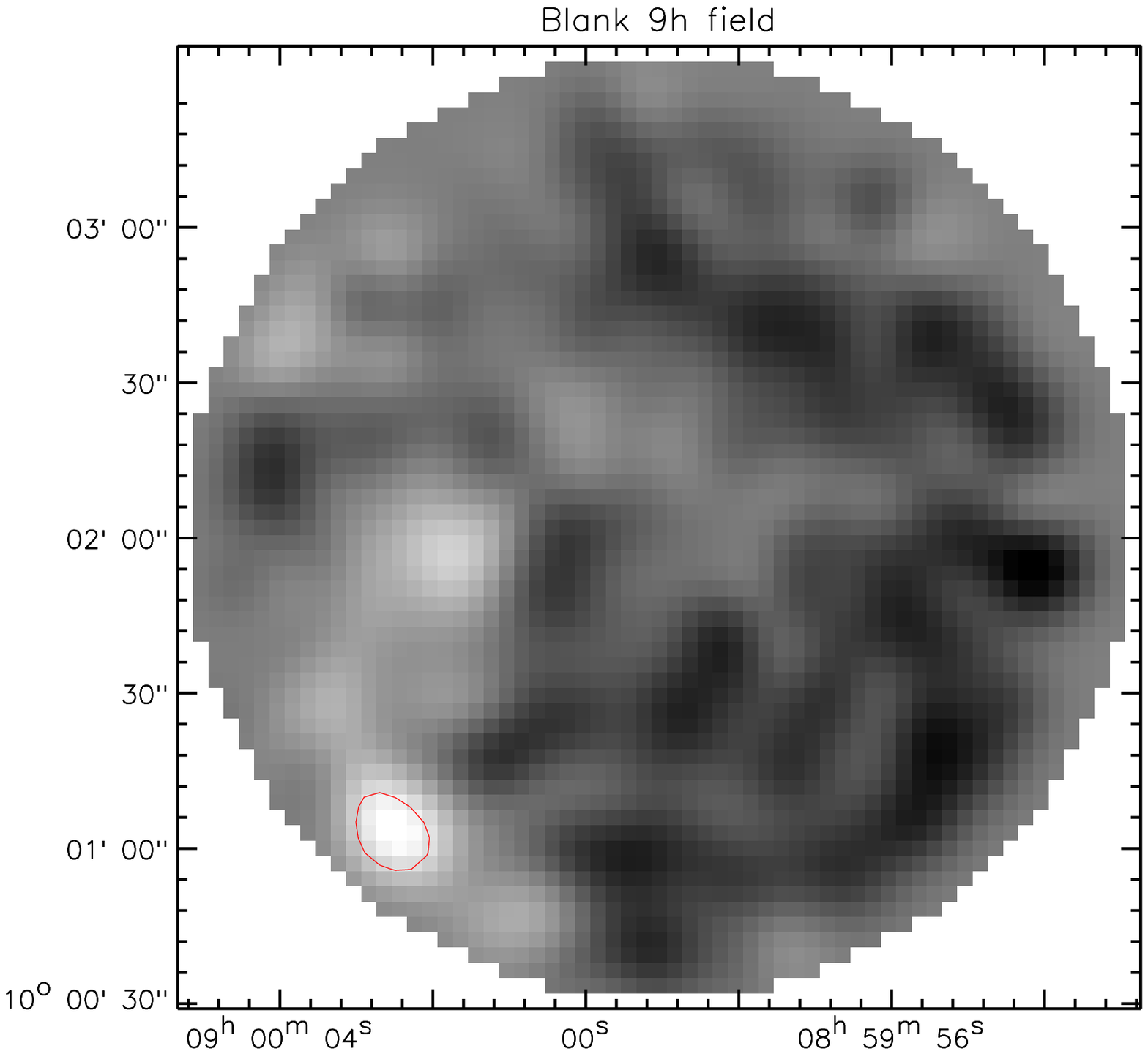,width=0.45\textwidth}
\epsfig{file=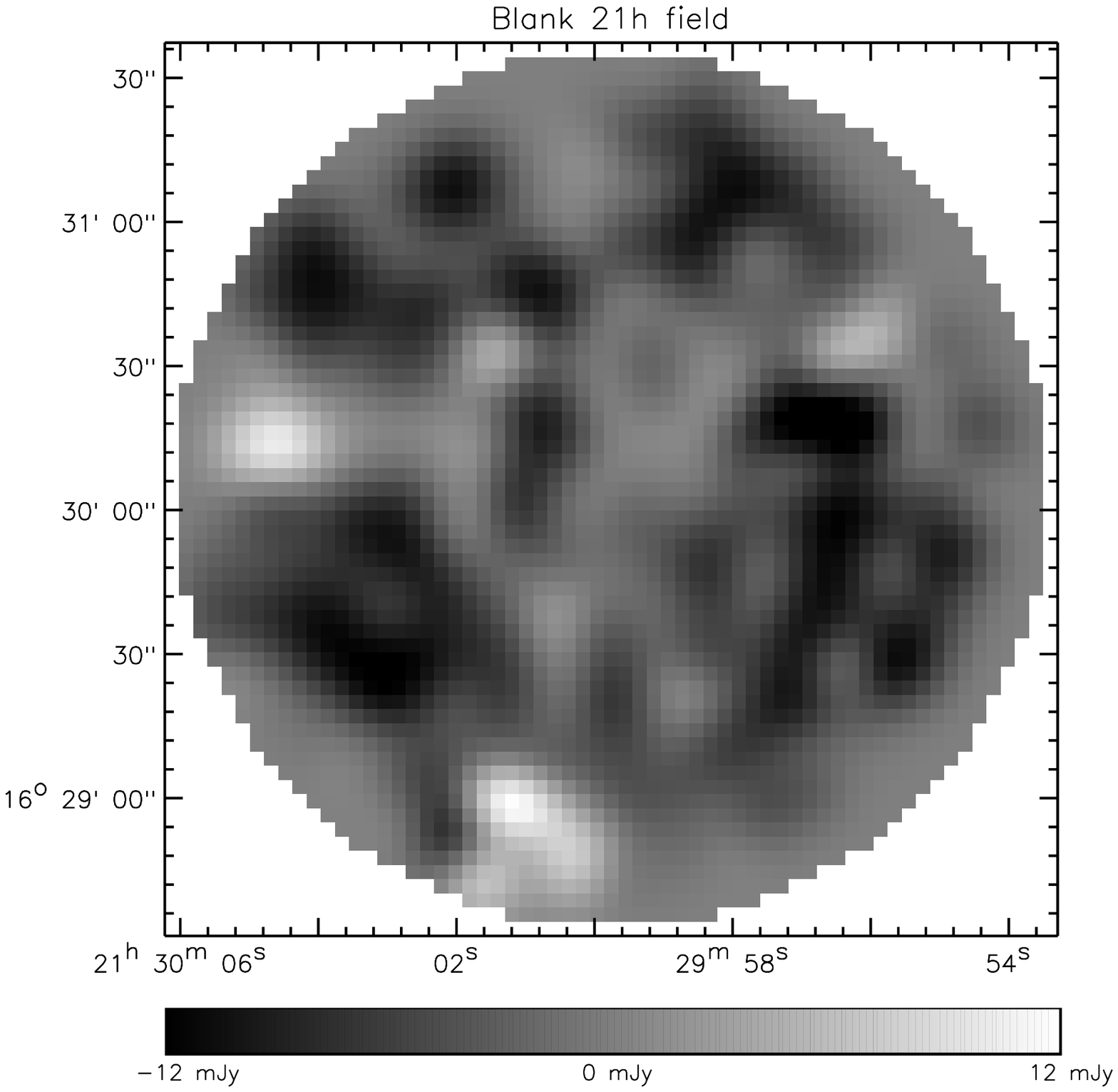,width=0.45\textwidth}
\caption{Maps of our target fields using a standard map-making
procedure (similar to the {\sc surf} map-making algorithm).  These
maps are not used at any point in our analysis, except to serve as a
check that large point sources are not present in the data.  Contours
show the point source cutoff levels in each field (quoted in the
text).  The 1$\sigma$ rms values of these maps are: 2.9 mJy in Cl$\,
0016$; 3.3 mJy in MS$\, 1054$; 3.6 mJy in Blank $09$; and 3.9 mJy in
Blank $21$.  The fluxes and (possible) locations of point sources
found and removed from later analysis are given in Table $2$.}
\label{fig:2}
\end{figure*}

Instead, the binned flux differences are convolved with a wavelet
similar to a `Mexican Hat' function:
\begin{equation}
\psi(\theta) = \psi_{0} \; \left(1 - \frac{\theta^{2}}{2
\sigma_{\mathrm{w}}^{2}} \right) \exp \left( \frac{- \theta^{2}}{2
\sigma_{\mathrm{w}}^{2}} \right).
\end{equation}
Here $\theta$ is an angular distance and $\sigma_{\mathrm{w}}$ is a
parameter charactering the wavelet's width. The normalization
$\psi_{0}$ is set such that the flux from a point source placed in a
blank field is returned identically at $\theta=0$. This wavelet is
more useful than the telescope PSF in this analysis because it returns
an amplitude which is insensitive to the average value in the map,
which is not expected to be zero in an SZ experiment.  The wavelet's
width was chosen to be $\sigma_{\mathrm{w}} = 6.3$ arcsec, slightly
smaller than the $\sigma$ corresponding to the FWHM of the telescope,
because this value maximizes the signal to noise ratio in simulated
measurements.  After convolution with this function, the positive and
negative extrema in the maps are the locations of possible point
sources in the source and reference beams respectively (Table 2).

\begin{table}
\caption{Candidate point sources which are removed.}
\label{tab:2}
\begin{tabular}{cccc}
\hline
Field       & Flux (mJy) & RA (2000) & DEC (2000) \\ \hline

Cl$\, 0016$ & $10$  & 0:18:32.7  & +16:25:18 \\

MS$\, 1054$ & $-11$ & 10:56:52.5 & -3:37:27  \\

MS$\, 1054$ & $-12$ & 10:56:53.0 & -3:37:45  \\

Blank $09$  & $13$  & 9:00:02.4  & +10:01:02 \\ \hline
\end{tabular}
\end{table}

The philosophy of this experiment's point source removal scheme is not
to develop a list of point sources \textit{per se}, but rather to
remove the flux contributed by any point-like source.  In the simple
method utilized here, possible point sources in a field are removed by
adopting a single flux level above which sources are found.  The
limiting flux level is estimated based on atmospheric extinction,
integration time and the noise equivalent flux density.  For each
field, we use the following thresholds: 9 mJy in Cl$\; 0016$, 9 mJy in
MS$\, 1054$; 11.5 mJy in Blank $09$; and 13 mJy in Blank $21$.  It is
found that the point source list matches that found in
\citet{Chapman2002} well.  The estimated variance given by this method
and the actual mean variance of the pixels agree to 10 per cent or
better in each field.

By determining the reference beam positions at which most of the
negative flux originates, it is found that the `two' negative point
sources in MS$\,1054$ are, in fact, probably caused by one bright
source.  The source appears to be two sources because the MS$\, 1054$
field was observed only at the beginning and the end of the Nov.~11
1999 shift.  The field rotated during the interval, thereby extending
the single source into `two' apparent sources.  Our best estimate of
this source's true position is 10:57:03.5 $\pm 2.0$ seconds in right
ascension and $-3$:39:00 $\pm 60$ arc seconds in declination.
\citet{Chapman2002} find another source of similar amplitude in this
cluster which did not happen to fall in any of our beams.  The
presence of these sources is consistent with the source count models
(e.g.~\citealt{Smail2002}).

The effect of greater than 3$\sigma$ positive sources is removed
directly from the time series using the measured amplitude, the
telescope pointing history and the telescope's PSF.  A negative
3$\sigma$ source is equally likely to be in either of the two
reference beam positions, so the same approach is not available.
Hence, data associated with detected negative sources are simply
excised from further analysis.

Monte Carlo simulations have been performed to develop an
understanding of the effects of point sources on this data.  These
simulations are fully explained in Zemcov, Newbury \& Halpern (2003)
(hereafter \citealt{Zemcov2003}), and the application of the results
found there are discussed in Sections $3.3$ and $5$.

\subsection{Model Fitting}

Because the SZ effect is extended on the sky, the simplest way to
isolate it in a cluster is to fit the data to a model.  Since the
signal to noise ratio in any given pixel is poor for this data set, a
simple isothermal $\beta$ model is adequate:
\begin{equation}
\Delta I^{\mathrm{D}}(\theta) = \Delta I_{0}^{\mathrm{D}} \; \left( 1
+ \frac{\theta^{2}}{\theta_{0}^{2}} \right)^{(1 - 3 \beta) / 2}.
\end{equation}
Here, $\Delta I_{0}^{\mathrm{D}}$ is a measure of the magnitude of the
increment at the centre of the cluster, $\theta$ is the angular
distance from the centre, and $\theta_{0}$ and $\beta$ are parameters
characterizing the cluster.  These parameters have already been
determined for these clusters from combinations of X-ray and SZ
decrement data (see Table $3$).  Although there is some uncertainty in
the values of $\beta$ and $\theta_{0}$, the error this causes in the
inferred $\Delta I_{0}^{\mathrm{D}}$ is negligible at this
experiment's level of precision.  We thus fix the angular dependence
and focus on determining the amplitude alone with our SCUBA data.

\begin{table}
\caption{Isothermal $\beta$ profile fit parameters.}
\label{tab:3}
\begin{tabular}{lllll}
\hline
Cluster	                & $\theta_{0}$ & $\beta$ &
$z$    & Reference           \\ \hline

Cl$\, 0016$               & $42.3$\arcs  & $0.749$ &
$0.55$ & \citet{Reese2000}   \\ 

MS$\, 1054^{\mathrm{a}}$  & $66.0$\arcs  & $0.75$  &
$0.83$ & \citet{Jeltema2001} \\

Blank $09$, $21$        & $50.0$\arcs  & $0.75$  &
       & Generic Values      \\
\hline
\multicolumn{5}{l}{$^{\mathrm{a}}$\citet{Jeltema2001} discuss the
difficulty of applying an} \\
\multicolumn{5}{l}{isothermal $\beta$ profile to MS$\, 1054$, which is a
double--cored} \\
\multicolumn{5}{l}{X-ray cluster.}
\end{tabular}
\end{table}

Because the SZ increment is non-zero at $450$\mums and the $450$\mums
array average is used to remove residual sky noise, the superscript
`D' is used as a reminder that the value quoted is actually a
difference between the $850$\mums signal and the $450$\mums array
average.  Retrieving the $850$\mums signal (denoted $\Delta
I_{0}^{850}$) from $\Delta I_{0}^{\mathrm{D}}$ will be discussed
further below.

Directly fitting the data to an isothermal $\beta$ model differenced
according to the telescope pointing history is the most
straightforward way to determine the peak SZ increment flux in a given
field.  Both the $\Delta I_{0}^{\mathrm{D}}$ and the 1$\sigma$ error
in $\Delta I_{0}^{\mathrm{D}}$ found from fitting to the differenced
isothermal $\beta$ model are listed in Table $4$.  The error in
$\Delta I_{0}^{\mathrm{D}}$ is found via the correlation matrix of the
linear fit.  $\sigma (\Delta I_{0}^{\mathrm{D}})$ is the square root
of the second diagonal element of this correlation matrix.  The fit
correlation matrix has very small off diagonal elements, meaning that
the fit is not correlated with the (meaningless) voltage offset value.

To test the hypothesis that much of the information about the SZ
intensity is contained in the difference $S_{b}^{850} (t)$, its
average is calculated over the array at each time step.  The mean of
these values should be greater than 0 and be detected with
approximately the same or slightly less significance as the detection
given by the fit to the isothermal $\beta$ model.  It is found that
the means of the Cl$\, 0016$ and MS$\, 1054$ fields are approximately
$1.5 \sigma$ and $2 \sigma$ above zero, respectively.  We conclude
that merely using the shape of the SZ distortion across the SCUBA
field of view underestimates its true amplitude.  

Note that intensities, $\Delta I_{\mathrm{SZE}}$, are generally quoted
in mJy per JCMT beamsize, which is related to temperature change via
the derivative of the blackbody distribution:
\begin{equation}
\Delta I_{\mathrm{SZE}} = 2 \, k_{\mathrm{B}}^{3} \, \Omega
\left( \frac{T_{\mathrm{CMB}}}{h c} \right)^{2} \frac{x^{4}
e^{x}}{(e^{x} - 1)^{2}} \Delta T_{\mathrm{SZE}}
\end{equation}
where $x = h \nu / k T_{\mathrm{CMB}}$ is the dimensionless frequency,
$T_{\mathrm{CMB}} = 2.725 \,$K, $\Delta T_{\mathrm{SZE}}$ is the
change in CMB temperature due to the SZ effect, and $\Omega = 5.74
\times 10^{-9}$ steradians per beamsize.

\begin{table}
\caption{SZ model fit results.}
\label{tab:4}
\begin{tabular}{lcc} \hline
Target      & $\Delta I_{0}^{\mathrm{D}}$ (mJy~beam$^{-1}$) & 
$\sigma (\Delta I_{0}^{\mathrm{D}})$ (mJy~beam$^{-1}$) \\ \hline

Cl$\, 0016$ & $1.5$  & $0.5$ \\

MS$\, 1054$ & $1.5$  & $0.7$ \\

Blank $09$  & $-0.4$ & $0.7$ \\

Blank $21$  & $1.1$  & $1.0$ \\ \hline
\end{tabular}
\end{table}

It is also desirable to make a crude `map' to check the radial profile
of the data.  A matrix inversion technique utilizes all of the
available data to make a map, including information about the
off-source pointings (e.g.~\citealt{Tegmark1995}, \citealt{Wright1996},
\citealt{Stompor2002}, \citealt{Hinshaw2003}).  This method involves
dividing the sky into a vector of pixels, creating matrices based on
pointing and flux data, and the inversion of a single matrix to create
a map. The matrix inversion itself is performed using a Singular Value
Decomposition (e.g.~\citealt{Press1992}).

Here, the pixel model of the sky is chosen to be the brightnesses of
annular rings centred on the cluster, as this conforms to the
approximately circular symmetry expected in the SZ signal. The rings
have $40$ arcsec widths to track the radial variation with reasonable
signal to noise ratio.  Once the fluxes of the annuli including their
errors and correlations have been estimated, they can be fit to the
isothermal $\beta$ model.  The fit is performed using both data and
model with a weighted mean of 0.  This ensures that the model's DC
term does not falsely increase the measured increment.  A generalized
$\chi^{2}$ statistic is determined, where $\mathbf{x}_{\mathrm{m}}$ is
a pixel's isothermal $\beta$ model value, and
$\mathbf{x}_{\mathrm{d}}$ is a pixel's data value. 
\begin{equation}
\chi^{2} = (\mathbf{x}_{\mathrm{m}} - \mathbf{x}_{\mathrm{d}})_{i} \
\mathbf{C}_{ij}^{-1} \ (\mathbf{x}_{\mathrm{m}} -
\mathbf{x}_{\mathrm{d}})_{j}.
\end{equation} 
Our observational strategy produces non-negligible correlations between
the annular pixels which are given in $\mathbf{C}_{ij}$, the
pixel--pixel covariance matrix.  Explicitly, $\mathbf{C}_{ij} = \left(
\mathbf{P}^{\mathrm{T}} \mathbf{N}^{-1} \mathbf{P} \right)^{-1}$,
where $\mathbf{P}$ is our pointing matrix and the noise is
$\mathbf{N}$.  $\mathbf{N}$ is assumed to be diagonal because removing
the fit to the $450 \,$\mums array average effectively acts as a
pre-whitening filter.  These correlations need to be carefully
included when the SZ amplitude is determined using this annular pixel
model and an isothermal $\beta$ profile.  Fig.~$3$ shows the annular
pixel fluxes as well as the best fit profile (Equation ($3$)) using
the parameters listed in Tables $3$ and $4$.  The two methods yield
very similar results, although we find that the direct fitting method
is more sensitive.

\begin{figure*}
\centering
\epsfig{file=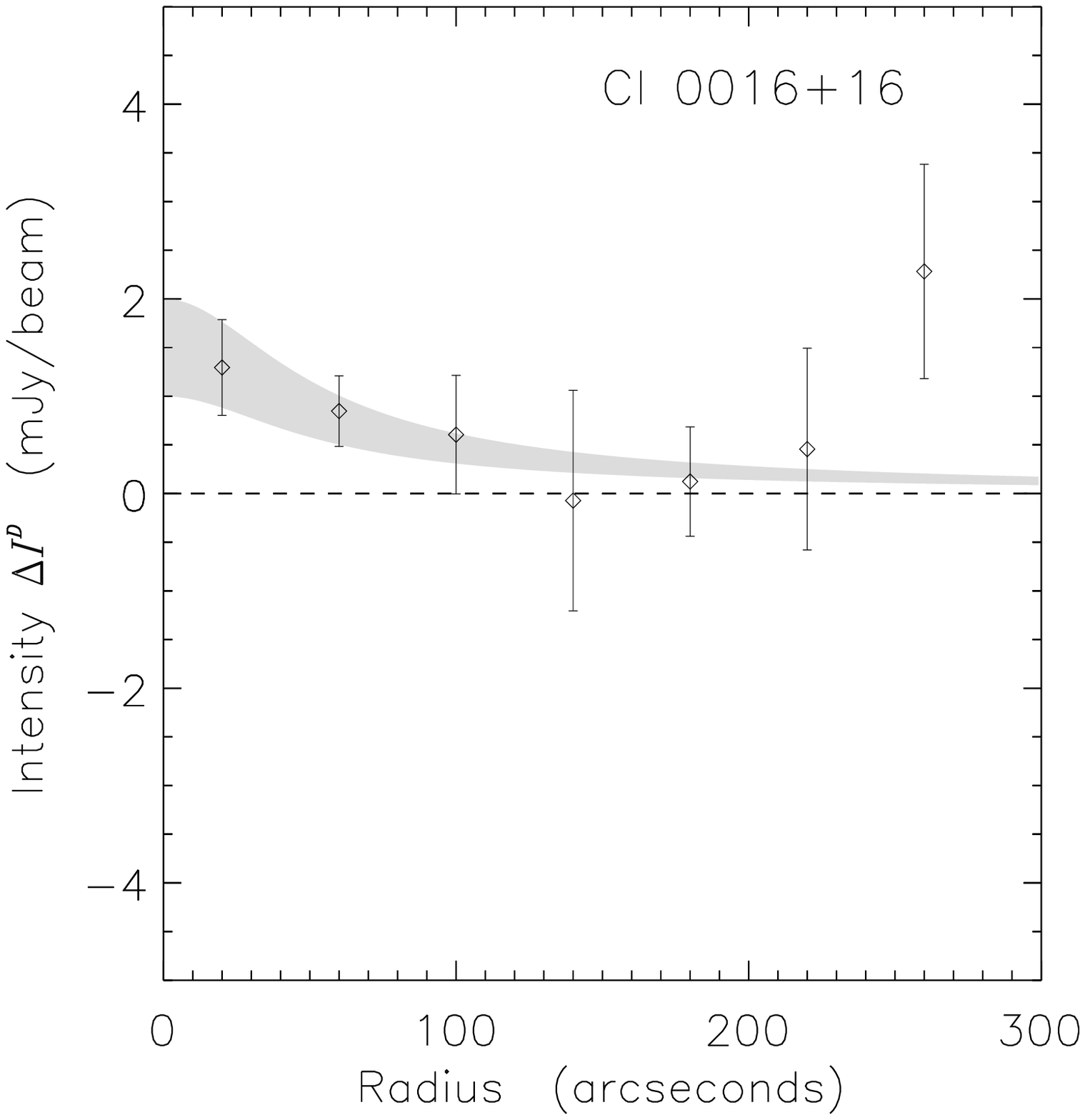,width=0.45\textwidth}
\epsfig{file=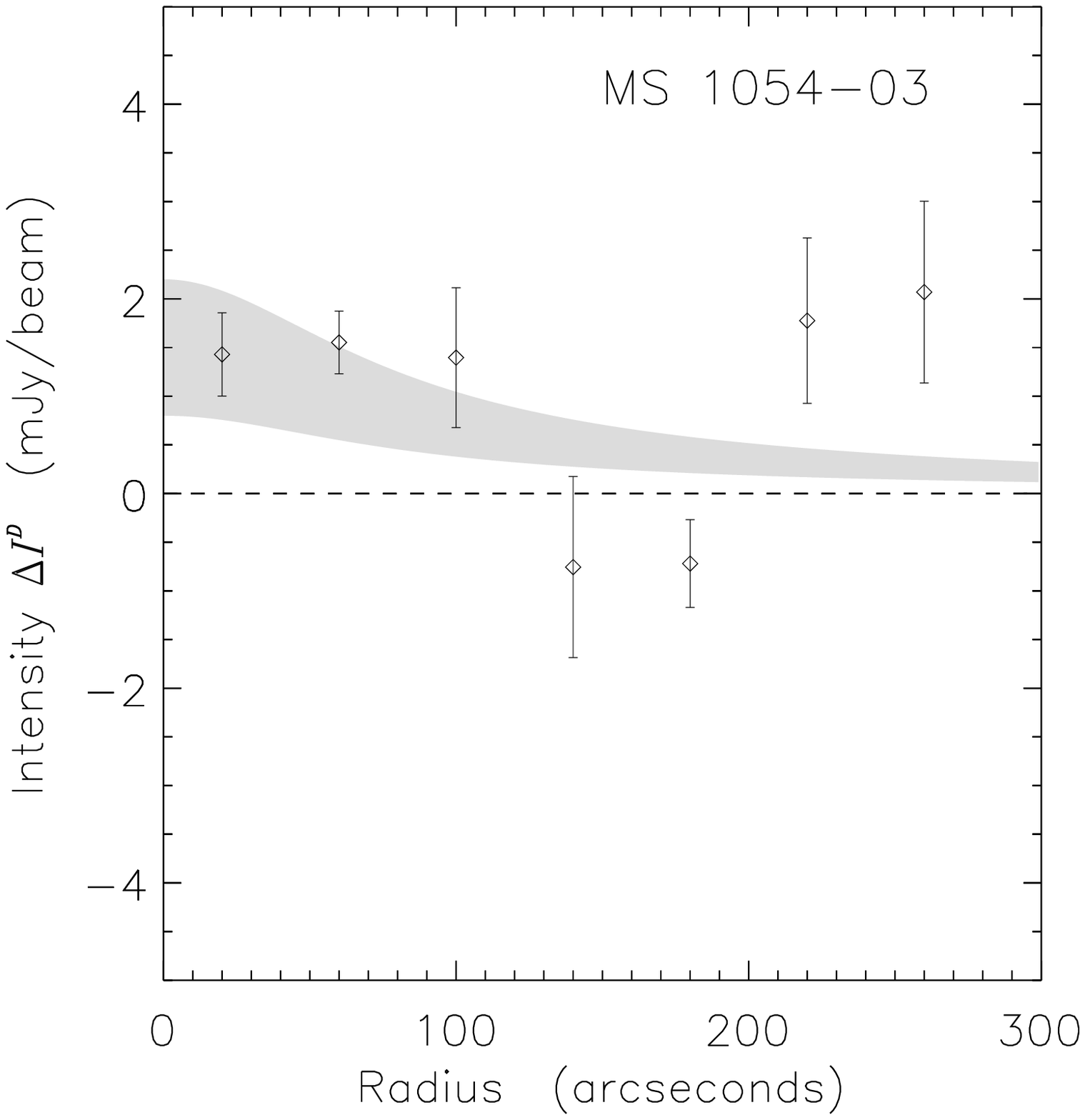,width=0.45\textwidth}
\epsfig{file=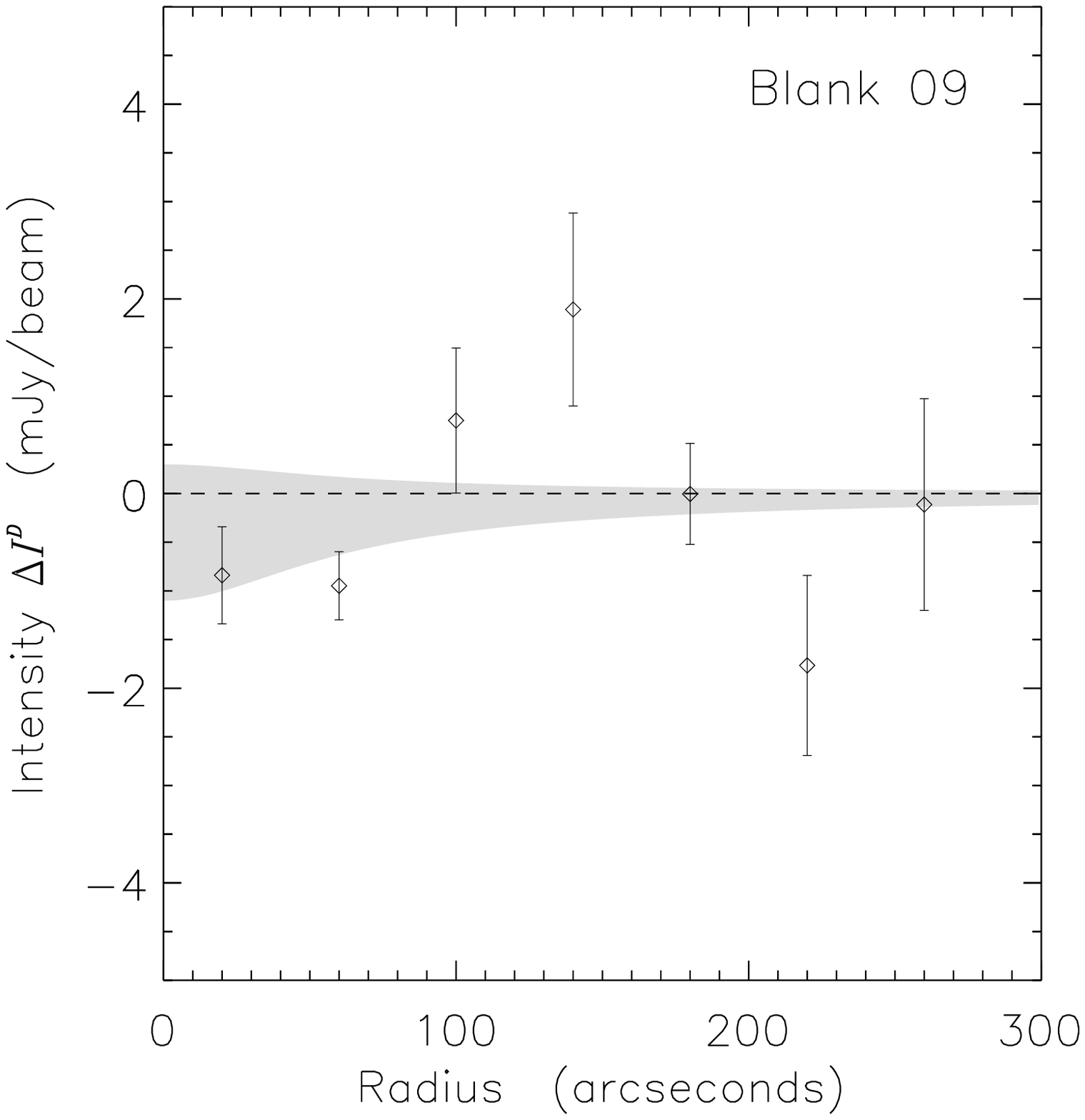,width=0.45\textwidth}
\epsfig{file=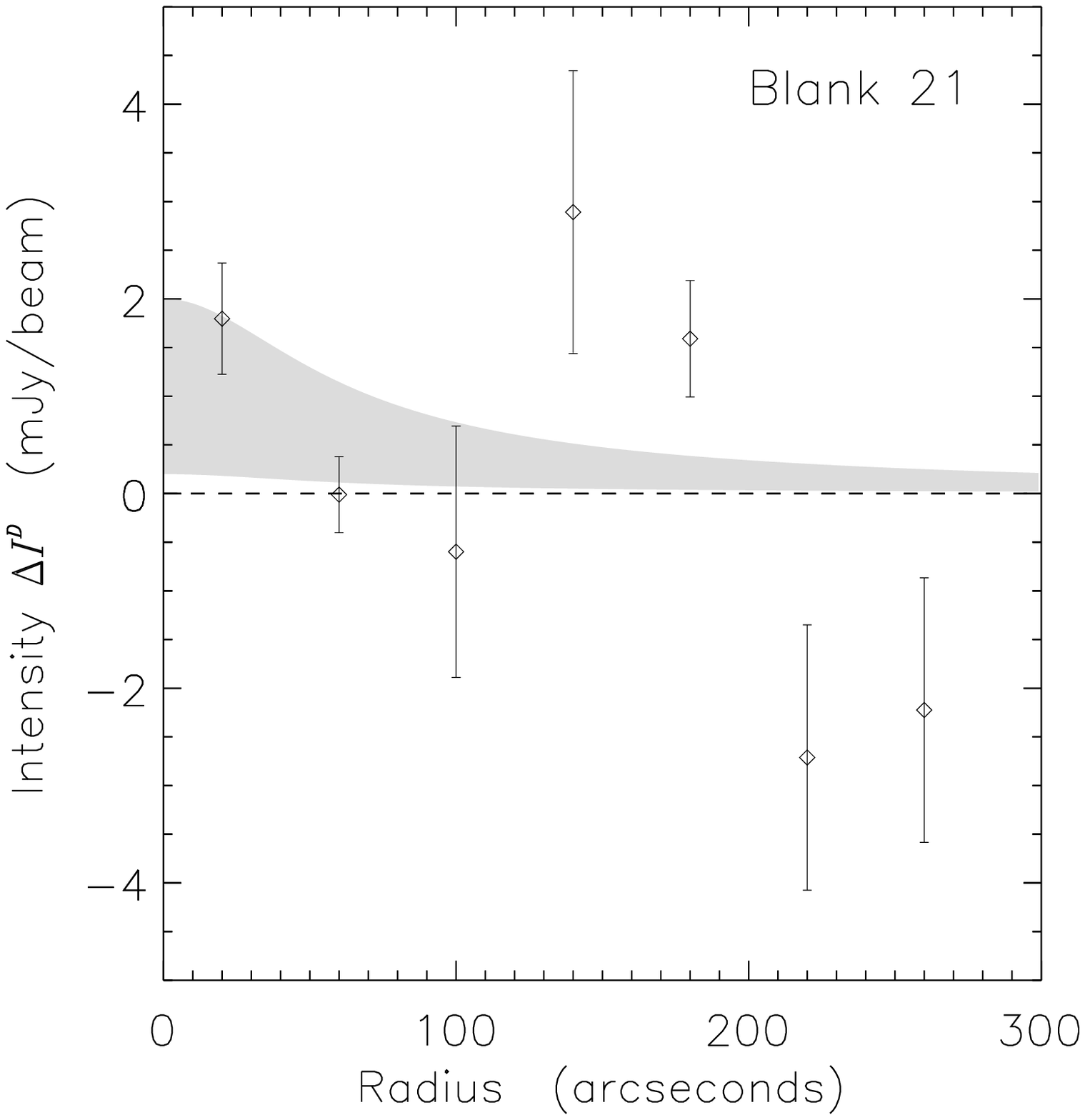,width=0.45\textwidth}
\caption{Data points determined via matrix inversion, together with
best fit isothermal $\beta$ profiles found from a direct fit for the
$4$ target fields shown as $\pm 1 \sigma$ confidence regions by the
grey band.  Note that the isothermal $\beta$ profiles are not best
fits to the data points shown but rather use the model in Table $3$
fit directly to the SCUBA differences.  The annular points are used to
assess the fields' radial behaviour and to determine the effect of
point sources in the beams.  As matrix inversion yields data with a
mean of zero, the mean of the best fit model is added to the data in
these plots.  The zero level is shown as a dashed line.}
\label{fig:3}
\end{figure*}

Generally, the figure of merit for the goodness of fit is that
$\chi^{2}$ should be approximately the number of degrees of freedom in
the data.  However, source confusion may have an effect on the
applicability of this figure of merit.  To test how well $\chi^{2}$
estimates the goodness of fit, Monte Carlo simulations similar to
those presented in \citet{Zemcov2003} are performed in order to
determine the probability that the cluster field fits are consistent
with the absence of an SZ increment.  The direct fit to the isothermal
$\beta$ model differences method is used to find $\Delta
I_{0}^{\mathrm{D}}$ in simulations containing realistic point sources
and with noise levels set to be the same as those in this experiment,
but with no SZ increment in the fields.  Point sources are removed at
the same level as in this experiment; Fig.~$4$ shows the results of
200 such simulations.  It is found that the fits associated with the
highest $10$ per cent of the $\chi^{2}$ statistics always yield drastically
incorrect increment values because of point source contamination.  We
therefore exclude the worst $10$ per cent of the fits.  Fig.~$4$ also
shows the SCUBA results for both clusters, each of which pass the
$\chi^{2}$ cutoff.  The results overlap, as the fits for these fields
give almost the same increment value.  If the SZ increment were absent
from Cl$\, 0016$ and MS$\, 1054$, the probability of obtaining the
$\Delta I_{0}^{\mathrm{D}}$ found is less than approximately $0.01$
for either cluster.

\begin{figure}
\centering
\epsfig{file=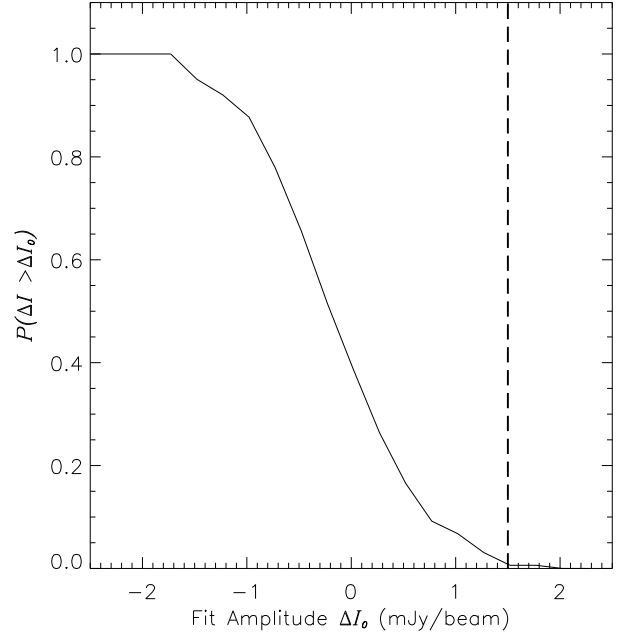,width=0.45\textwidth}
\caption{Results of simulations to determine the effect of confusion
on these measurements.  This plot shows the cumulative probability of
finding an increment value $\Delta I_{0}$ if no increment were
present, $P(\Delta I \! > \! \Delta I_{0})$, as a function of fit
amplitude $\Delta I_{0}$ found from fitting to the differenced
isothermal $\beta$ model.  The simulations use realistic noise
including point sources and no SZ effect.  Point source removal has
been implemented (see Section $3.2$ for details).  The values
determined for the two clusters in this study are shown by the dashed
vertical line.  This function is not symmetric about zero because
gravitational lensing preferentially brightens positive point sources,
thereby biasing the point source removal scheme.}
\label{fig:4}
\end{figure}

The $\chi^{2}$ values found for Cl$\, 0016$ are $3.7$ for $6$ degrees
of freedom via the seven pixel fit method, and a $\chi^{2}$ of
$178402$ for $178406$ degrees of freedom using the direct fit to model
differences.  We conclude that the fits are fully consistent with the
presence of an increment in this cluster.

Our confidence in the parameters found for MS$\, 1054$ is less than
that for Cl$\, 0016$.  Based on its $\chi^{2}$ of $13.4$ for $6$
degrees of freedom from fitting the annular pixels to an isothermal
$\beta$ model, this cluster would normally be flagged as a poor fit.
The fit may be poor because the isothermal $\beta$ model does not
describe the distribution of MS$\, 1054$'s intracluster electron gas
particularly well \citep{Jeltema2001}.  However, the fit may become
tolerable if the nature of the double--cored electron gas distribution
is taken into account.  Nevertheless, the reduced $\chi^{2}$ from
fitting to the isothermal $\beta$ model differences is $258438$ for
$258204$ degrees of freedom in this cluster.  The reduced $\chi^{2}$
statistic associated with this fit falls well below the worst 10 per
cent of reduced $\chi^{2}$ statistics in the Monte Carlo simulations
discussed above.  The fit is therefore acceptable based on this
criterion.

The parameters determined for the blank fields are also given in Table
$4$.  From the fits to annular pixels, the Blank $09$ field has a
$\chi^{2}$ of $11.7$ for $6$ degrees of freedom and the Blank $21$
field has a $\chi^{2}$ of $13.6$ for $6$ degrees of freedom.  The
$\chi^{2}$ from fitting to the model differences are poor in both
fields.  Both fields also seem to exhibit unremoved, confused point
sources.  The simulations of \citet{Zemcov2003} show that fields with
no SZ increment form a well-defined locus on the $I_{0}$--$\chi^{2}$
plane, and both of the blank fields in this experiment would be inside
the confused regime and hence be rejected as SZ increment candidates
based on their $\chi^{2}$.

For the purposes of comparison, the line $\Delta I_{0} = 0 \,$mJy
beam$^{-1}$ is fit to the data using the direct fit method.  The
$\Delta \chi^{2} (= \chi^{2} - \chi^{2}_{\mathrm{min}})$ values for
these fits are 6.7 for Cl$\, 0016$, 6.6 for MS$\, 1054$, 0.5 for Blank
$09$ and 2.0 for Blank $21$.  These values are in good agreement with
the significance of the detections.

Although the MS$\, 0451$ and Blank $22$ fields were found to be
unsatisfactory for this experiment earlier in our analysis, this
analysis pipeline also rejects these fields.  Both fields have $\Delta
I_{0}$ consistent with $0 \,$mJy per beam, and poor $\chi^{2}$ due to
the number and location of sources within them.  This is an objective
check that retrieving an SZ increment from data heavily polluted by
point sources is difficult.

\section{Individual Cluster Analysis \& Results}

In this section our data are used to determine the $850$\mums SZ
increment values in Cl$\, 0016$ and MS$\, 1054$.  A maximum likelihood
method is used to determine the best fit Compton $y$ parameter (which
parametrizes the integrated pressure in the cluster along the line of
sight) and, in Cl$\, 0016$, the peculiar velocity of the cluster along
the line of sight.

\subsection{Cl$\, \mathbf{0016}$+$\mathbf{16}$ $\mathbf{\Delta I_{0}^{850}}$
Determination}

\citet{Worrall2003} find that Cl$\, 0016$ has $T_{\mathrm{e}} \simeq 9
\, $keV, implying that it has a relativistic electron population which
may give a significant correction near the positive peak of the SZ
spectrum.  Therefore, the corrections found by \citet{Itoh1998}, which
are good to fifth order in $k_{\mathrm{B}} T_{\mathrm{e}} /
m_{\mathrm{e}} c^{2}$, are used to determine the correct shape of the
thermal SZ distortion.  These corrections, which represent a
$10$--$20$ per cent change to the flux at $850$\mums for standard
cluster parameters, are applied to all of the spectral calculations,
and are implicit in the use of the term `thermal' SZ effect below.

Because the SCUBA filters have a finite bandwidth, we need to
determine the effective central frequencies of this measurement.  This
calculation must be performed because the SZ spectrum differs from the
grey-body spectra SCUBA usually observes.  To do this, the `equivalent
frequency' is defined as:
\begin{equation}
\bar{\nu} = \frac{\int_{0}^{\infty} \nu F( \nu ) S( \nu ) d
\nu}{\int_{0}^{\infty} F( \nu ) S( \nu ) d \nu},
\end{equation}
where $F( \nu )$ is the actual SCUBA filter's response and $S ( \nu )$
is the SZ effect spectrum over the region of interest.  The results of
this calculation are listed in Table $5$; we conclude that the
frequency shift from the nominal values are negligibly small.

\begin{table}
\caption{Equivalent $\nu$, $\lambda$ for SZ spectra.}
\label{tab:5}
\begin{tabular}{llll}
\hline
Filter        & $\bar{\nu}$    & $c / \bar{\nu}$ \\ \hline

$850 \,$\mums & $347 \,$GHz   & $865 \,$\mums      \\

$450 \,$\mums & $652 \,$GHz   & $460 \,$\mums      \\ \hline
\end{tabular}
\end{table}

Foreground dust emission could also contaminate the measurement of the
SZ increment in these clusters.  \textit{ISO}\footnote{We have used
archival data from the Infrared Space Observatory (\textit{ISO}), a
European Space Agency mission with the participation of ISAS and NASA
\citep{Kessler1996}.} data are used to estimate the dust contribution.
The telescope pointings are used to sample the $180$\mums and $90$\mums
\textit{ISO} map of the cluster region in exactly the same way as the SCUBA
data.  Errors for each measurement are determined from the error map
and incorporated into the simulated time stream.  These data are then
fit to the model differences.  The $\chi^2$ values imply that the fits
are very poor; this is expected because any dust emission in the
region (Galactic cirrus, extragalactic point sources, or the cluster
itself) is unlikely to be distributed like the isothermal $\beta$
model.  However, the dust may still contribute to the SZ signal and so
must be removed.  The dust contribution is determined by fitting a
modified black body of the form:
\begin{equation}
I_{\mathrm{d}}(\nu,T_{\mathrm{d}}) = \alpha \frac{\nu^{3 +
\gamma}}{e^{h \nu / k_{\mathrm{B}} T_{\mathrm{d}}} - 1}.
\end{equation}
Here, the parameters are the spectral index $\gamma$, dust temperature
$T_{\mathrm{d}}$, and normalization $\alpha$, which is determined
using both of the \textit{ISO} data points weighted by their errors.
$\gamma$ is taken to be $2.0$ and $T_{\mathrm{d}}$ to be $20 \,$K,
parameters which are typical for \textit{ISO}--detected cirrus at
these wavelengths (e.g. \citealt{Lamarre1998}).  Using this model, the
dust emission at $450$\mums is determined to be $I_{\mathrm{d}}(660 \,
\mathrm{GHz}) = 0.07 \pm 0.04$ mJy~beam$^{-1}$ for Cl$\, 0016$.  This
value is not very sensitive to variations in the parameters $\gamma$
and $T_{\mathrm{d}}$.  The fit is corrected for dust emission by
the addition of this value, which is a few per cent correction.

Because our analysis scheme subtracts the $450 \,$\mums flux, a method
of retrieving $\Delta I_{0}^{850}$ from $\Delta I_{0}^{\mathrm{D}}$ is
required.  The actual quantity subtracted from each $850$\mums
bolometer's time stream is the average of the $450$\mums array at each
time step.  This average contains the SZ signal sampled as described
in Fig.~1, with $\Delta I_{0}^{450}$ being the peak isothermal $\beta$
model amplitude.  The average value of the $450 \,$\mums array is
determined by sampling the model SZ profile in the same way as in this
experiment.  A map is made based on this data, and the average of this
map in the field of view is constructed.  It is found that there is a linear
relationship between the input SZ amplitude and the average of the
differenced map, given by $\Delta I_{\mathrm{Ave}}^{450} = A
\Delta I_{0}^{450}$, where $A$ is a linear coefficient and $\Delta
I_{\mathrm{Ave}}^{450}$ is the average of the difference map over the
field of view.  This means that $\Delta I_{0}^{850}$ can be found via:
\begin{equation}
\Delta I_{0}^{850} = \Delta I_{0}^{D} + A \Delta I_{0}^{450},
\end{equation}
where $A$ depends only on the shape of the isothermal $\beta$ model.
$A=0.72 $ in Cl$\, 0016$, including the gain difference between the
$850$\mums and $450$\mums arrays.

To determine the best fit Compton $y$ parameter and peculiar velocity
$v_{\mathrm{pec}}$, we utilize a likelihood function which searches
the $y$ - $v_{\mathrm{pec}}$ parameter space for its maximum.
Previous decrement measurements at $15 \, $GHz \citep{Grainge2002},
$20 \, $GHz \citep{Hughes1998} and $30 \, $GHz \citep{Reese2000} are
used alone to determine a best-fit $y$ value, and also in conjunction
with our measurement to help constrain both the thermal and kinetic SZ
effects in Cl$\, 0016$.  Alone, these values yield $y = (2.38 \pm
0.12) \times 10^{-4}$ in this cluster, assuming $v_{\mathrm{pec}} = 0
\,$km s$^{-1}$.  Our value alone yields $y = (2.2 \pm 0.7) \times
10^{-4}$ under the same assumption.  This measurement is therefore
consistent with earlier results.

We find that the $850 \,$\mums SZ increment value for Cl$\, 0016$ is
$\Delta I_{0}^{850} = 2.2 \pm 0.7$ mJy~beam$^{-1}$.  This corresponds
to $0.38 \pm 0.12$ MJy sr$^{-1}$ or $\Delta T_{\mathrm{CMB}} = 1.2 \pm
0.4$ mK in thermodynamic temperature units.  Fig.~$5$ shows the
corrected $850$\mums data point, along with the radio and \textit{ISO}
results.  Also plotted are the best fit thermal SZ effect and dust
emission spectra, and the position of the $450$\mums band--centre for
reference.

\begin{figure}
\centering
\epsfig{file=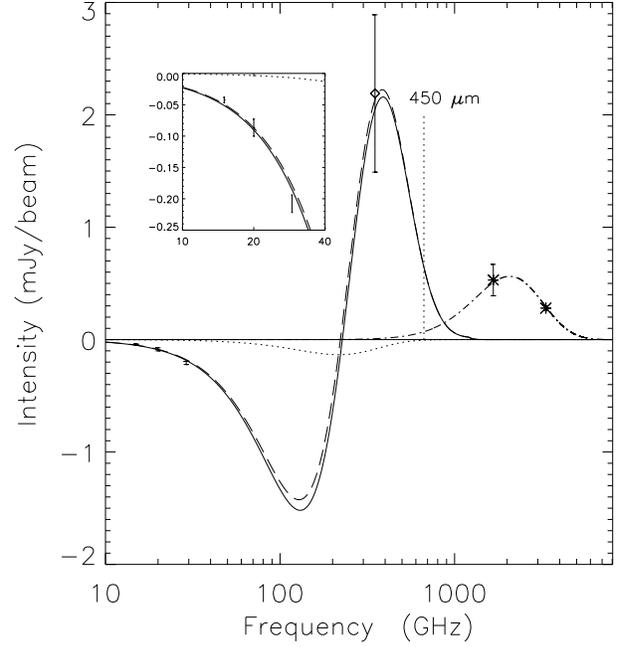,width=0.45\textwidth}
\caption{Sources of extended emission in the Cl$\, 0016$ field.
The dashed line shows the corrected thermal SZ effect determined from
both our $850$\mums point and low frequency decrement measurements.
The inset shows the low frequency data.  The dot-dash line shows the
dust emission determined from \textit{ISO} measurements (asterisks).
The best fit kinetic effect is shown as a dotted line.  The sum of the
thermal and kinetic effects is shown as a solid line.  For reference,
the position of $450$\mums (dotted vertical line) and zero flux
(horizontal solid line) are shown.  Our $850$\mums point is shown as a
diamond, and has been corrected for both the $450$\mums subtraction
and dust emission.}
\label{fig:5}
\end{figure}

In addition to constraining $y$, the amplitude of the kinetic SZ
effect is also constrained by this likelihood analysis.  Using both
our measurement and the three decrement measurements, it is found that
the marginalized best fit parameters are $y = (2.38 \pm
_{0.34}^{0.36}) \times 10^{-4}$ and $v_{\mathrm{pec}} = 400 \pm
_{1400}^{1900}$ km~s$^{-1}$ in Cl$\, 0016$.  Assuming
$v_{\mathrm{pec}} = 0$, these points yield $y = (2.38 \pm 0.11) \times
10^{-4}$.  Fig.~$5$ also shows the inferred kinetic SZ effect from the
$850$\mums data.  This result is therefore only a weak constraint on
the line of sight peculiar velocity.

A kinetic SZ effect in Cl$\, 0016$ may be mimicked by a primary
fluctuation in the CMB, because the spectral distributions of the two
effects are the same. However, the primary CMB anisotropies at the
scales of this experiment, $\ell > 5000$, are about $1 \mu \,$K rms
(\citealt{Hu2002}, \citealt{Borys1999}). This corresponds to a $\pm
0.005 \,$mJy beam$^{-1}$ signal, which is negligible at our level of
precision.

\subsection{MS$\mathbf{\, 1054-03}$ $\Delta I_{0}^{850}$ Determination}

The value of $y$ for MS$\, 1054$ is found by determining the
difference $I^{850} - A I^{450}$ for the SZ distortions defined by
$1.0 \times 10^{-5} \leq y \leq 1.0 \times 10^{-3}$, where $A = 0.81$
for this cluster.  The $y$ values giving differences most closely
matching the data and 1$\sigma$ error points are the best fitting $y$
parameter and error bars.  This method yields $y = \left( 2.0 \pm 1.0
\right) \times 10^{-4}$ for this cluster, including relativistic
corrections.  These $y$ values correspond to an increment of $\Delta
I_{0}^{850} = 2.0 \pm 1.0 \,$mJy~beam$^{-1}$, or $\Delta
T_{\mathrm{CMB}} = 1.0 \pm 0.5 \,$mK.

Because little other relevant data exist for this cluster, further
analysis is not possible.  The dust contamination cannot be estimated
because there are no \textit{ISO} data, and the available dust maps
\citep{Schlegel1998}, which are of poor angular resolution, provide little
spectral information.  However, it appears that the dust contamination
is at a similar level to that found in Cl$\, 0016$, and hence is
probably negligible.  Although one published estimate of the thermal
decrement magnitude already exists for MS$\, 1054$ \citep{Joy2001}, a
worthwhile estimate of its kinetic effect cannot be made because no
explicit $y$ parameter or $\Delta T_{\mathrm{CMB}}$ is available.

\section{Discussion}

In principle, measurements of the SZ increment, in combination with
lower frequency measurements, can yield a great deal of information.
For example, the line of sight peculiar velocity and possibly even the
cluster temperature can be determined via measurement of the SZ
increment.  However, measurement of the SZ increment with SCUBA is
difficult.  There are three essential components which combined to
make this experiment a success.

The first component is a carefully planned observational strategy
designed to maximize SCUBA's sensitivity to an SZ signal.  Using a
large chop throw to reduce the tendency to subtract SZ signal and
chopping in azimuth so that sky rotation reduces the effect of point
source contamination are both parts of this.  Also, using the
$450$\mums array average to remove the spurious sky signal, thereby
keeping SZ information on scales larger than the array size, maximizes
the available signal in the data.  Controls such as the blank fields,
and to a lesser extent the reflector data, are required to check for
systematic effects in both the instrument and analysis methods.

The second necessary component of this experiment is the use of
custom-designed software to handle the data analysis, which is
particularly important for fitting the differenced data directly to a
model.

The third important component of this analysis is the removal of point
sources in order to reduce their contaminating effect.  Because sub-mm
bright, high $z$ background sources are relatively common in cluster
fields, observations of the SZ increment could suffer prohibitively
from the effects of point sources.  The basic principle we employ is
that point sources affect only a few pixels and hence can be
distinguished from an extended SZ effect.  Our data have a confusion
limit of approximately $0.8 \,$mJy~beam$^{-1}$ for SZ signals.
Integration to SCUBA's fundamental confusion limit of $\simeq 0.5
\,$mJy beam$^{-1}$ rms (which would require weeks of integration time)
gives an SZ confusion of $\sim 0.5 \,$mJy beam$^{-1}$.  Our
simulations show negligible reduction in the confusion limit of these
observations when point sources with amplitude less than about 5 mJy
are removed \citep{Zemcov2003}.  This limit would require
approximately 2 shifts of integration time.  Unfortunately, decreasing
the confusion limit floor below $0.5$ mJy beam$^{-1}$ would require
higher angular resolution observations.  A corollary to this is that
SZ increment detections from other experiments with lower resolution
than SCUBA, such as SuZIE and ACBAR, may be confusion dominated if
unsupported by higher resolution data.

These measurements confirm the SZ effect in 2 galaxy clusters at an
amplitude consistent with the measured decrements.  A weak constraint
on the kinetic effect in Cl$\, 0016$ is found which is consistent with
the result in \citet{Benson2003}.  We quote a marginal 2$\sigma$ SZ
increment detection in MS$\, 1054$, but cannot do much more with the
data at this point.

It is clear that large reductions in the uncertainty of the value for
the kinetic effect velocity in Cl$\, 0016$ will not be achievable
because of source confusion with this type of instrument.  This means
that SCUBA is not capable of determining the amplitude of the kinetic
effect in any specific cluster, given the expected level of
SZ--derived peculiar velocities (e.g. \citealt{Sheth2001}).  It is
likely that the only way to use SCUBA for constraining peculiar
velocities is through a statistical survey of many clusters.

Nevertheless, we have demonstrated that, with sufficient care, it is
possible to measure the $850$\mums increment at the JCMT.  Such SCUBA
observations of the SZ effect at $850$\mums can be combined with
measurements at other frequencies to study individual clusters in more
detail.  Follow--up observations with high resolution sub-mm
instruments are likely to become an important part of forthcoming SZ
cluster surveys.

\section*{Acknowledgments}

We are grateful to Dr.~Mark Birkinshaw for his helpful comments.  This
work was supported by the Natural Sciences and Engineering Research
Council of Canada.  The James Clerk Maxwell Telescope is operated by
The Joint Astronomy Centre on behalf of the Particle Physics and
Astronomy Research Council of the United Kingdom, the Netherlands
Organization for Scientific Research, and the National Research
Council of Canada.  We would like to acknowledge the staff at JCMT for
facilitating these observations.  This research has made use of NASA's
Astrophysics Data System, the SIMBAD database, operated at CDS,
Strasbourg, France, and the Canadian Astronomy Data Centre.  E.~P.~was
funded by NSERC and NASA grant NAG5-11489 during the course of this 
work.

\parskip 10pt 
\noindent This paper has been produced using the Royal Astronomical
Society/Blackwell Science \LaTeX\ style file.

\end{document}